\documentclass[twocolumn,showpacs,floatfix,prl,footinbib]{revtex4}

\usepackage{amssymb}
\usepackage{graphicx}
\usepackage{dcolumn}
\usepackage{bm}

\begin{document}

%\draft
\title{Many Body Effects on the 
Transport Properties of Single-Molecule Devices}
\author{P. S. Cornaglia}
\author{H. Ness}
\author{D. R. Grempel}
\affiliation{CEA-Saclay, DSM/DRECAM/SPCSI, B\^at. 462, F-91191 Gif sur Yvette,
France}

\begin{abstract}
The conductance through a molecular device including electron-electron 
and electron-phonon interactions is calculated using 
the Numerical Renormalization Group method.
At low temperatures and weak electron-phonon coupling 
the properties of the conductance can be explained in terms of 
the standard Kondo model with renormalized parameters.
At large electron-phonon coupling a charge analog of the Kondo effect
takes place that can be mapped into an anisotropic Kondo
model. In this regime the molecule is strongly polarized by a gate
voltage which leads to rectification in
the current-voltage characteristics of the molecular junction.
\end{abstract}


\maketitle

Electronic conduction through nanoscale systems connected to
external electrodes exhibits a number of special 
features
that are of great importance
when one considers their potential 
as electronic devices.

The transport properties of such systems are strongly affected  by 
({\it e-e}) and electron-phonon ({\it e-ph}) interactions. 
For example, Coulomb blockade effects have been shown to dominate
the transport in quantum dots \cite{CB_Qdots} and even in single
molecules weakly coupled to the electrodes \cite{JPark_2002,Kubatkin_2003}. 
The Kondo effect has been observed in quantum dots \cite{Gordon1998} 
and single molecules having well defined spin and charge 
states \cite{JPark_2002,WLiang_2002,Yu_2004}.
Furthermore the effects of {\it e-ph} coupling have been 
observed in Inelastic Electron Tunneling Spectra of 
small molecules adsorbed on surfaces \cite{WHoXX}
and in molecular-scale transistors made of ${\rm C}_{60}$ 
molecules \cite{HPark_2000}. Finally, 
 {\it e-ph} coupling is also known to play an important role in 
the transport through organic conjugated molecules in which 
transport is mediated via polaron or soliton propagation \cite{cond_polym}.

It is usually assumed that energy scales for {\it e-e} interaction
are much larger than those for {\it e-ph} interaction.
However, it has been recently shown that Coulomb charging energies 
of single molecules can be considerably reduced by screening due to 
the electrodes \cite{Kubatkin_2003}.
These energies can be of the order of a few 100 meV 
(instead of a few eV for an isolated molecule) and therefore are 
of the same order of magnitude of the relaxation energies induced
by {\it e-ph} coupling in the same systems.
Interesting physics then arises when the energy scales of 
{\it e-e} and {\it e-ph} interactions merge.
By studying such cases, insights on the origin of the features 
in the conductance can be brought 
and the influence of the {\it e-ph} coupling on the Kondo effect
can be addressed.

Although there have been many theoretical investigations of electron 
transport in the presence of either {\it e-e} or {\it e-ph}
interactions, 
the case in which both interactions are  
present has only been 
studied recently~\cite{Mitra_2003,Kuo_2002,Flensberg_2003} either in the
 high temperature or weak {\it e-ph} coupling regimes.

In this Letter, we report results of non-pertubative calculations 
of the linear transport properties through a molecular device  
including {\it e-e} and {\it e-ph} interactions.
The calculations are performed using the Numerical Renormalization 
Group (NRG) technique for a broad range of characteristic parameters.

We study a model of a molecule with a single relevant electronic level 
coupled linearly to a vibrational mode of frequency $\omega_0$ and
to the left (L) and right (R) metallic electrodes. 
The Hamiltonian of the system is
\begin{equation}
H=H_{M} + H_{E} + H_{M-E},
\label{hamil}
\end{equation}
where the first two terms describe the isolated molecule 
 and electrodes, respectively, and the last term describes their 
coupling. We have 
\begin{eqnarray}
H_{M}&=&\varepsilon_d n_d + U n_{d \uparrow}n_{d \downarrow}-\lambda\;
\left(n_d-1 \right)\left(a^{}+a^{\dagger}\right)\\
\nonumber
&&+\omega_0a^{\dagger}a^{}\;,\\
H_{E}&=&\sum_{{\bf k},\sigma,\alpha=\{{\rm L,R}\}} \varepsilon_{\alpha}({\bf k})\;
c^{\dagger}_{{\bf k} \sigma \alpha} c^{}_{{\bf k} \sigma
\alpha}\;%\left(\alpha={\rm L,R}\right)\;
,\\
H_{M-E}&=&\sum_{{\bf k},\sigma,\alpha}\;V_{{\bf k}\alpha}\left(d^{\dagger}_{\sigma}
\;c^{}_{{\bf k}\sigma\alpha} + c^{\dagger}_{{\bf
k}\sigma\alpha}\;d^{}_\sigma \right)\;.
\end{eqnarray}
Here, $n_d = \sum_{\sigma} d^{\dagger}_\sigma d^{}_\sigma$ is the
charge of the molecule, $\varepsilon_d$ is the position of the electronic
molecular level relative to the Fermi level 
of the electrodes and $U$ is 
the Coulomb repulsion between two electrons that occupy the same molecular 
level. $a^\dag$ creates a phonon of frequency $\omega_0$ and
$\lambda$ is the {\it e-ph} coupling constant.
We consider for
simplicity the case of identical electrodes and
contacts.  
In the wide-band limit the conductance $G$ of the
molecular junction  at zero bias is~\cite{Meir1992,Jauho1994} 
\begin{equation}
\label{conductance}
G=\left.{dI\over dV}\right|_{V=0}={2e^2\over h}\ \pi \Delta
 \int_{-\infty}^{\infty} d\omega
 \left(-{\partial f(\omega)\over \partial \omega}\right)
 \rho_{d}({\omega})\;, 
\end{equation}
where $\rho_{d}(\omega)=-\pi^{-1}{\rm Im} G_{dd}(\omega)$, 
$ G_{dd}(\omega)$ is the exact electronic Green function of the
molecule in the presence of the leads, $f(\omega)$ the Fermi distribution
 and
$\Delta = 2 \pi \rho_0 \langle V^2_{\bf k} \rangle_{\rm FS}$. 
Here, $\rho_0$ is the electrodes' electronic density of states at the Fermi
level and the brackets denote an average over the Fermi surface.  
Hamiltonian (\ref{hamil}) is electron-hole 
symmetric for all values of $\lambda$ at
$\varepsilon_d=\varepsilon_d^\star=-U/2$ where $\langle
n_d\rangle=1$. 
This point we will
referred to as the {\it symmetric point} in the following. 

At zero temperature Eq.~(\ref{conductance}) reads
\begin{equation}
\label{conductance-T0}
G=G_0\pi \Delta \rho_d(0)=G_0 \sin^2\left(\pi {\langle n_d
 \rangle/ 2}\right)\;,  
\end{equation} 
where $G_0=2e^2/h$ is the quantum of conductance and  
the second equality follows from Luttinger's 
theorem~\cite{Hewson-book}. 
 At $T=0$ the conductance $G$ thus takes 
its maximum value $G_0$ at the symmetric point.

Before presenting the results of a full numerical solution of the
problem we discuss  
two limiting cases that can be treated analytically 
 providing us with a qualitative picture of the dependence of
$G$ upon $T$ and $\varepsilon_d$. 

The Hamiltonian $H_M$ of the isolated 
molecule can be readily diagonalized. The eigenfunctions are direct
products of electronic states (denoted by a subscript $e$)
 and oscillator states. The eigenstates and eigenvalues are
%\begin{eqnarray} 
%\begin{array}{ll} 
%\left |0,m\right> = \tilde{U}^{-}\left |0\right>_e
%\left|m\right>, & E_{0,m}=-\frac{\lambda^2}{ \omega_0} + 
% m\omega_0,\\
%\left |\sigma,m\right> = \left |\sigma\right>_e\left|m\right>,
% & E_{\sigma,m}=\varepsilon_d 
%+ m\omega_0,\\
%\left |2,m\right> = \tilde{U}^{+}\left |\uparrow \downarrow\right>_e\left|m\right>, & E_{2,m}=-{\lambda^2\over \omega_0} + 2\varepsilon_d + U\\
%&\;\;\;\;\;\;\;\,\;\;\;\;\;+ m\omega_0,
%\end{array}
%\end{eqnarray}    
\begin{eqnarray}\label{eq:enes}
\begin{array}{ll} 
\left |0,m\right> = \tilde{U}^{-}\left |0\right>_e
\left|m\right>\!, & E^0_m=-\frac{\lambda^2}{ \omega_0} + 
 m\omega_0,\\
\left |\sigma,m\right> = \left |\sigma\right>_e\left|m\right>\!,
 & E^\sigma_m=\varepsilon_d 
+ m\omega_0,\\
\left |2,m\right> = \tilde{U}^{+}\left |\uparrow \downarrow\right>_e\left|m\right>\!, & E^2_m=-{\lambda^2\over \omega_0} + 2\varepsilon_d + U + m\omega_0,
\end{array}
\end{eqnarray}    
where $\tilde{U}^{\pm} =\exp{\left[\pm{\lambda/\omega_0}
\left(a^{\dagger}-a\right)\right]}$ and  $\left|m\right>$
is the $m$-th excited state of the harmonic oscillator.
Two limiting cases can be considered.

\noindent {\it Weak electron-phonon
coupling}, $2\lambda^2/\omega_0 \ll U$. In this case   
the ground-state of the isolated
molecule is the spin-doublet 
$\left|\sigma,0\right>$. 
There is a large charge excitation gap
 $\sim U_{\rm eff} = U - 2 \lambda^2/\omega_0$ at
the symmetric point and  
the low-energy excitations of the full system are spin fluctuations 
described by the usual 
Kondo Hamiltonian~\cite{Hewson-book}. Using standard 
second-order perturbation theory the coupling constant can be expressed 
in terms of a single matrix element: 
$\langle \sigma,0|H_{M-E}|0,m\rangle \langle 0,m
 |H_{M-E}|\sigma,0\rangle\;= 
\langle \sigma,0 |H_{M-E}|0,m\rangle \langle 0,m
 |H_{M-E}|-\sigma,0
 \rangle\;
 \propto\;|\langle 0 |\tilde{U}^+|m\rangle|^2$. At 
$\varepsilon_d=\varepsilon_d^\star$ we find the Kondo coupling constant ~\cite{Stephan} 
\begin{equation}
\label{jk-wc}
J_K(\lambda)\ \rho_0 \approx {8\Delta \over \pi U}\;\sum_{m=0}^\infty
{|\langle 0 | \tilde{U}^{+}| m \rangle|^2\over 1-
 {2\lambda^2 \over \omega_0 U}+ {2 m \omega_0 \over U}}\;,
\end{equation}
where $|\langle 0 | \tilde{U}^{+}| m \rangle|^2=
e^{-(\lambda/\omega_0)^2}(\lambda/\omega_0)^{2m}/m!.$

Below the Kondo temperature
 $T_K \propto \exp\left[-1/(J_K\rho_0)\right]$, 
 $G(\varepsilon_d^\star)~\sim~G_0$.
Expanding Eq.~(\ref{jk-wc}) around $\lambda=0$ 
we find $J_K/J_K(0) \approx 1+2 \left[1 + U/(2\omega_0)\right]^{-1}
\lambda^2/(U\omega_0)$. 
The {\it e-ph} coupling thus leads to an {\it increase} of the Kondo 
temperature in this regime.
For $T > T_K$, Coulomb blockade peaks separated by $U_{\rm eff}$ 
are expected at gate voltages 
 $\varepsilon_d \approx -\lambda^2/\omega_0$
and $\varepsilon_d \approx -U + \lambda^2/\omega_0$. 

Expanding  Eq.~(\ref{conductance-T0}) around the symmetric point 
we have $G(\varepsilon_d)-G(\varepsilon_d^\star) \sim - G_0\; (\chi_c\pi/2)^2
\left(\varepsilon_d - \varepsilon_d^\star\right)^2$, 
where $\chi_c $ is the charge susceptibility, inversely proportional to
the charge gap. 
The width  of the $T=0$ conductance peak, 
$\Delta \varepsilon_d \approx \chi^{-1}_c \approx U_{\rm eff}$, 
{\it decreases} with increasing $\lambda$ in the weak coupling 
regime.

\noindent {\it Strong electron-phonon coupling},
 $2\lambda^2/\omega_0 \gg U$. In this regime the ground-state
 doublet of the isolated molecule is composed of the states 
$\left|2,0\right>$ and
 $\left|0,0\right>$, degenerate at the symmetric point. 
The low-energy excitations of the full system are now {\it charge} 
fluctuations 
and there is a large gap for spin fluctuations.  
The low-energy excitations are described by  
 an effective Kondo model in which the role of the spin is played by
 a pseudo-spin
variable that represents the two states of the lowest lying doublet. 
{\it A priori} there is no rotational invariance in pseudo-spin
 space and, in fact, in the effective Hamiltonian two different matrix elements appear:   
$\langle 2,0 |H_{M-E}|\sigma,m\rangle \langle \sigma,m |H_{M-E}|2,0\rangle\;
 \propto\;|\langle 0 |\tilde{U}^+|m\rangle|^2$, and 
$\langle 2,0 |H_{M-E}|\sigma,m\rangle \langle \sigma,m |H_{M-E}|0,0\rangle\;
\propto
 \langle 0|\tilde{U}^-|m\rangle\langle m|\tilde{U}^+|0\rangle \equiv 
(-1)^m |\langle 0 |\tilde{U}^+|m\rangle|^2$.
The effective model is thus the {\it anisotropic} Kondo model (AKM) with   
couplings $J_\parallel$ and $J_\perp$ given by
\begin{equation}
J_{\left(\parallel,\perp\right)}\; \rho_0 \approx {8\Delta \over \pi
U}\sum_{m=0}^\infty 
 (\pm 1)^m {|\langle 0 | \tilde{U}^{+}| m \rangle|^2 \over 
 {2\lambda^2 \over \omega_0 U} -1 + {2 m \omega_0 \over U}},\nonumber
\end{equation}
and  Kondo temperature $T_{\rm AKM}$ given by~\cite{Costi1996}: 
\begin{equation}
T_{\rm AKM} \propto D \left({J_\perp / J_\parallel}\right)
^{1\over J_\parallel \rho_0}.
\label{takm}
\end{equation} 
 Asymptotically, 
$J_\perp/J_\parallel \approx
\exp\left[-2(\lambda/\omega_0)^2\right]$ and $T_{ AKM} \approx \exp\left[-
\alpha (\lambda/\omega_0)^4\right]$ with
$\alpha=\pi\omega_0/\Delta$. 
In this regime the Kondo temperature {\it decreases} sharply with 
increasing $\lambda$.
In contrast to the weak {\it e-ph} coupling case, no Coulomb blockade
peaks are expected for $T>T_{\rm AKM}$.
Another important difference from the weak $\lambda$ case is that, now, 
the charge susceptibility ($\propto T^{-1}_{\rm AKM}$) is very large. 
Then, the width of the conductance peak versus gate voltage 
$\Delta \varepsilon_d \approx T_{AKM}$
also decreases sharply as $\lambda$ increases. 
The region in the $T-\varepsilon_d$ plane where $G\approx G_0$ is thus 
strongly suppressed by a strong {\it e-ph} coupling.

We now turn to the discussion of our numerical results obtained using the 
NRG method~\cite{Wilson1975,Krishnamurthy1980}.
The original NRG method has been modified to include {\it e-ph} 
coupling~\cite{Hewson_2002} and to calculate the spectral density
$\rho_d(\omega)$ according to Ref.~\cite{Hofstetter2000}. 
We have performed calculations for a broad range of parameters.
We set $k_B=1$ and take the half-bandwidth of the electrodes as the 
unit of energy.
In the following, we present results for the set of parameters
 $\Delta=0.016$, $\omega_0=0.05$ and $U=0.1$. 

\begin{figure}[tbp]
\includegraphics[width=8.5cm,clip=true]{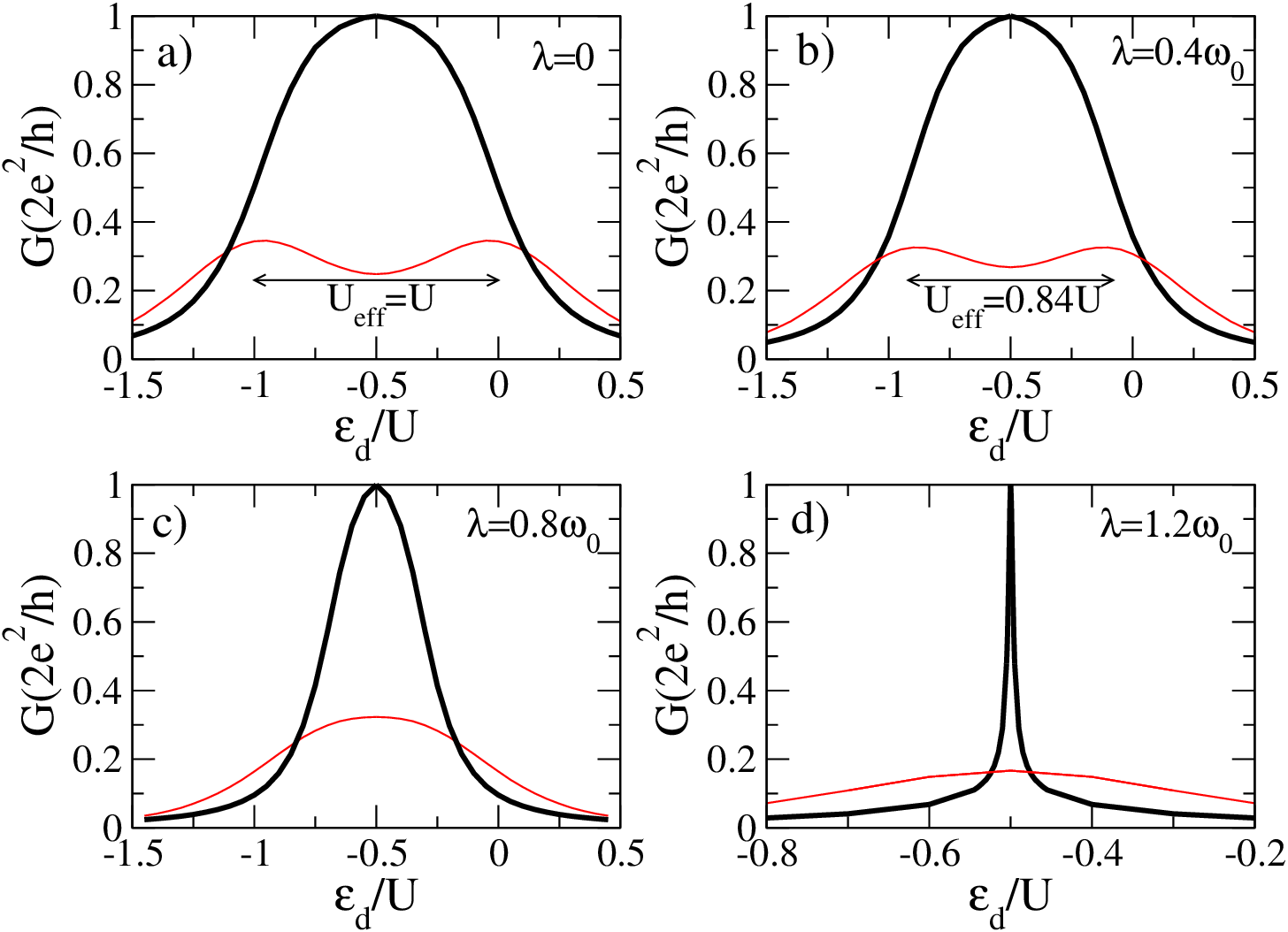}
\caption{Conductance $G$ versus gate voltage
 $\varepsilon_d$ for different values of $\lambda$ (a-d).
(thick lines: $T=0$, thin lines: $T=0.016$).
Parameters are $U=0.1$, $\Delta=0.016$ and $\omega_0=0.05$.}
\label{fig:fig1}
\end{figure}

Figure~\ref{fig:fig1} shows the linear conductance
$G$ as a function of the gate voltage 
$\varepsilon_d$ for $T=0$ and $T=\Delta$ 
and several values of the {\it e-ph} coupling $\lambda$. 
For $\lambda=0$ [Fig.\ref{fig:fig1}(a)] and $T=\Delta$  the   
 Coulomb blockade peaks~\cite{Glazman1988,Ng1988}
 separated by the charging energy $U$ are
clearly seen. With decreasing $T$  the
conductance in the Coulomb blockade valley
increases as the Kondo effect develops. 
At zero temperature $G$ has a single 
peak of width 
$\Delta \varepsilon_d \sim U$ centered at $\varepsilon_d^\star$.  
The main effect of a weak
coupling to the vibrational mode  at finite  $T$
[Fig.\ref{fig:fig1}(b)] is a reduction of the distance between
the Coulomb blockade peaks, now given by 
$U_{\rm eff}$ as anticipated above.  
The width of the $T=0$ peak is also of the order of 
$U_{\rm eff}$.  This large width is a consequence of the rigidity of the
 ground state of the molecule against charge fluctuations.

For strong {\it e-ph} coupling  [Fig.\ref{fig:fig1}(d)] the
features change qualitatively. A single peak is observed
in the conductance {\it at all temperatures} and its width sharply 
decreases with $T$. The peak narrowing  
results from the dramatic increase of the charge susceptibility 
of the molecule in the strong coupling
regime. 
Moving away from the symmetric point 
by application of a gate voltage produces an effect 
similar to that of applying a magnetic field in the standard
Kondo effect~\cite{Costi2001}:
the degeneracy of the ground-state doublet is broken and
the Kondo resonance in the spectral density is 
destroyed~\footnote{The charge Kondo effect is robust in
the presence of a magnetic field. 
No Kondo peak splitting and no suppression of the 
conductance is expected to occur upon application of a magnetic field 
in the strong {\it e-ph} coupling regime.}.

For $U_{\rm eff}\alt \Delta$ [Fig.\ref{fig:fig1}(c)]
the system is in a mixed valence regime in which the four charge
states of the molecule 
are nearly degenerate for $\varepsilon_d\sim\varepsilon^\star_d$. 
A single peak is then observed in the conductance at all temperatures
with a  $T=0$ width that is determined by the hybridization $\Delta$.

In all regimes discussed so far, the $T=0$ conductance at
$\varepsilon_d^\star$ is perfect 
as required by Eq.~(\ref{conductance-T0}). 

\begin{figure}[tbp]
\includegraphics[width=8.5cm,clip=true]{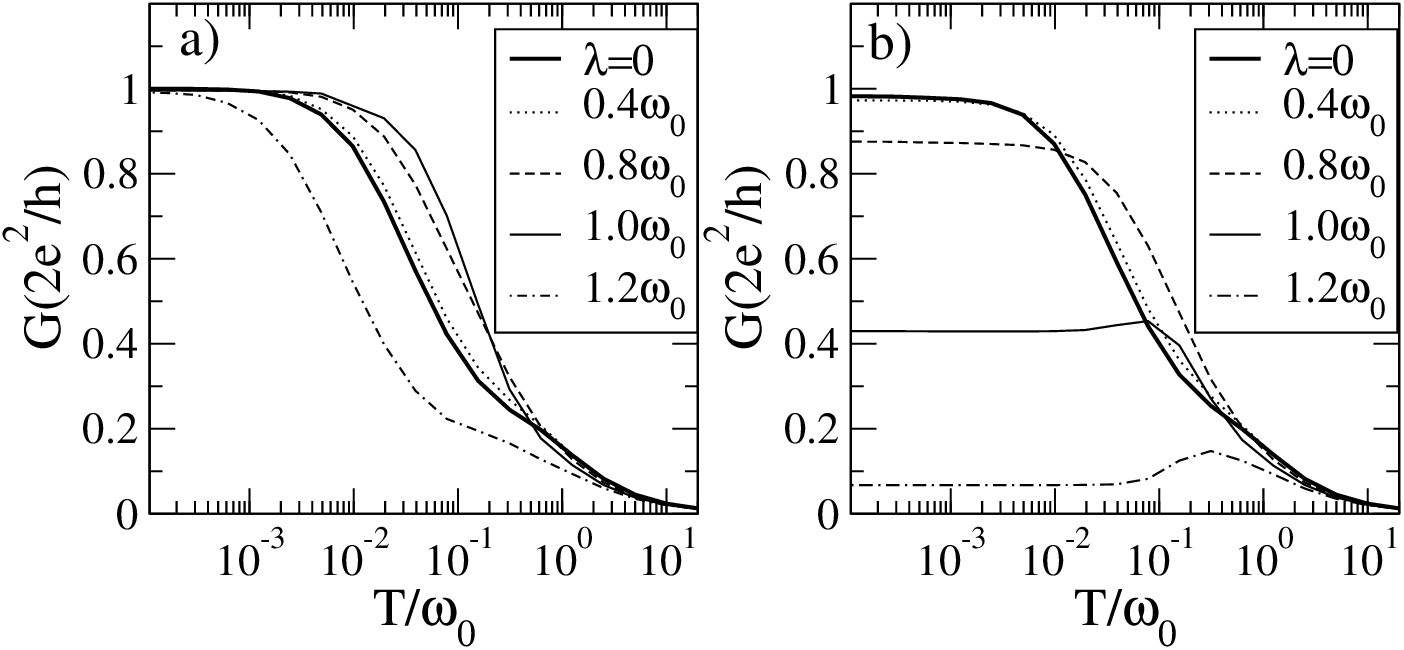}
\caption{Zero-bias conductance $G$ versus temperature $T$ for different 
values of $\lambda$. (a) Symmetric point 
$\varepsilon_d=-U/2=-0.05$. (b) Asymmetric case, 
$\delta\varepsilon_d=\varepsilon_d-\varepsilon_d^\star=0.01$. 
Other parameters as in Fig. \ref{fig:fig1}.} 
\label{fig:GdT}
\end{figure}
Figure \ref{fig:GdT} shows the $T$-dependence of $G$ 
  for several values of $\lambda$.
At the symmetric point [Fig.\ref{fig:GdT}(a)] $G\sim G_0$ below a  
temperature scale that depends on the coupling. 
This scale varies non-monotonically with $\lambda$, first increasing  
 and then decreasing. 
For very small $\lambda$ the behavior in the asymmetric case
 [Fig.\ref{fig:GdT}(b)] is similar to 
the previous one. For the larger couplings, however,  
 $G\ll G_0$ at all temperatures. 
\begin{figure}[tbp]
\includegraphics[width=7.5cm,clip=true]{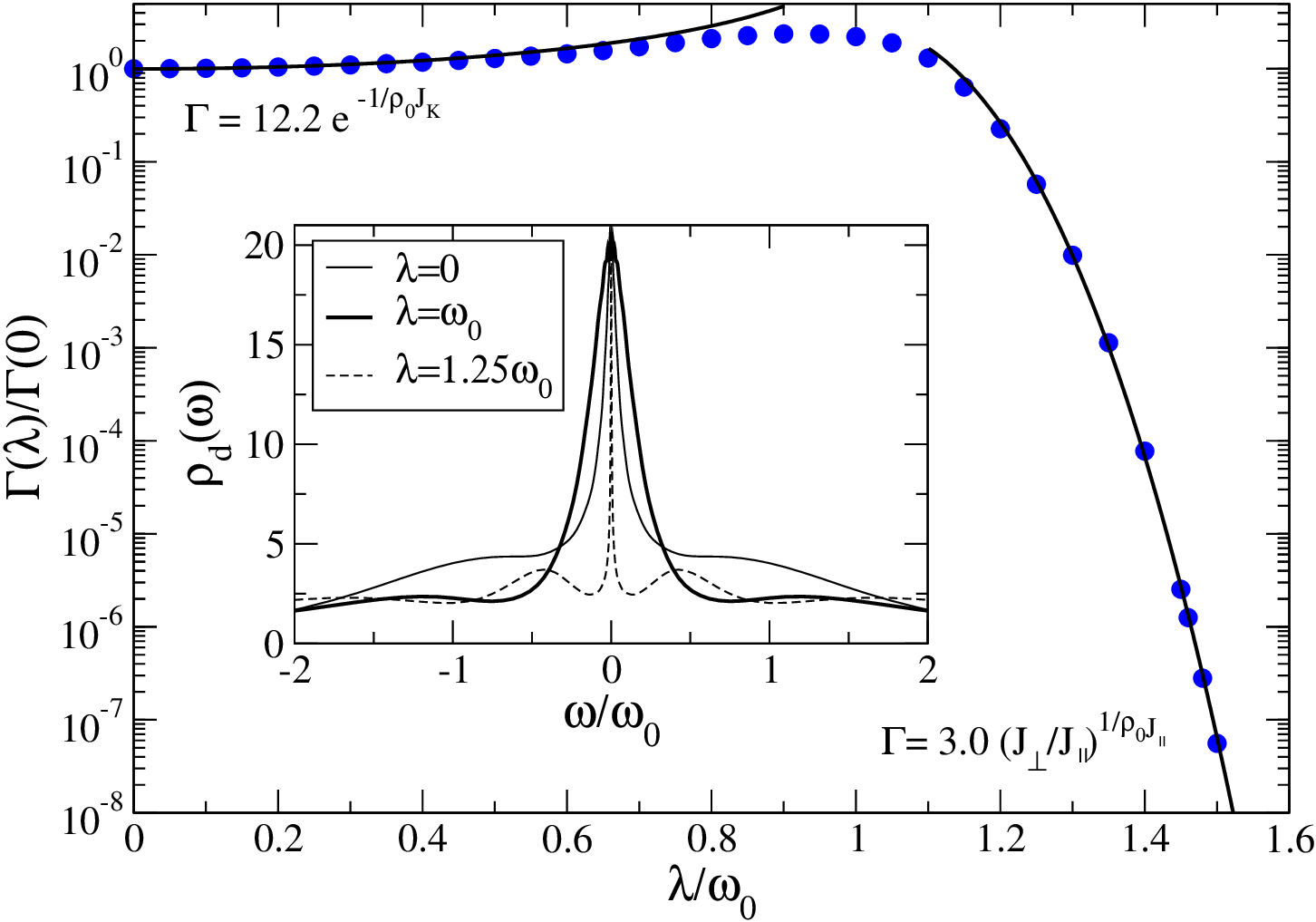}
\caption{Full width at half maximum $\Gamma$ of the central peak in the spectral density 
$\rho_d$ versus $\lambda$ at the symmetric point 
$\varepsilon_d=\varepsilon_d^\star$. 
Solid lines are fits based on the given analytical expressions (see
text). Inset: $\rho_d(\omega)$ for three values of $\lambda$.
Other parameters as in Fig. \ref{fig:fig1}.}
\label{fig:TKl}
\end{figure}

The inset in Fig.~\ref{fig:TKl} shows the spectral density
 $\rho_d(\omega)$ at the symmetric point for three values of the
{\it e-ph} coupling. 
The width of the central peak
 is renormalized by the interaction.
It is also non-monotonic as  a function of $\lambda$ 
as shown in the main plot where 
fits to the theoretical expressions discussed above are also shown.
For weak coupling region 
we use $\Gamma = A\;\exp\left[-1/(J_K\rho_0)\right]$ with $J_K$ given by
Eq.(\ref{jk-wc}). 
For strong coupling 
we use $\Gamma = A'\;T_{\rm AKM}$ with $T_{\rm AKM}$ given by
Eq.(\ref{takm}), instead. 
Only the amplitudes $A, A'$ are fitting parameters. 
It can be seen from Fig.~\ref{fig:TKl} that the agreement with the
predicted widths $\Gamma(\lambda)$ is excellent.
For intermediate couplings 
the charge fluctuations are no longer blocked by the Coulomb interaction 
and the system is in the mixed valence regime where $\Gamma={\cal O}(\Delta)$.
\begin{figure}[tbp]
\includegraphics[width=7.5cm,clip=true]{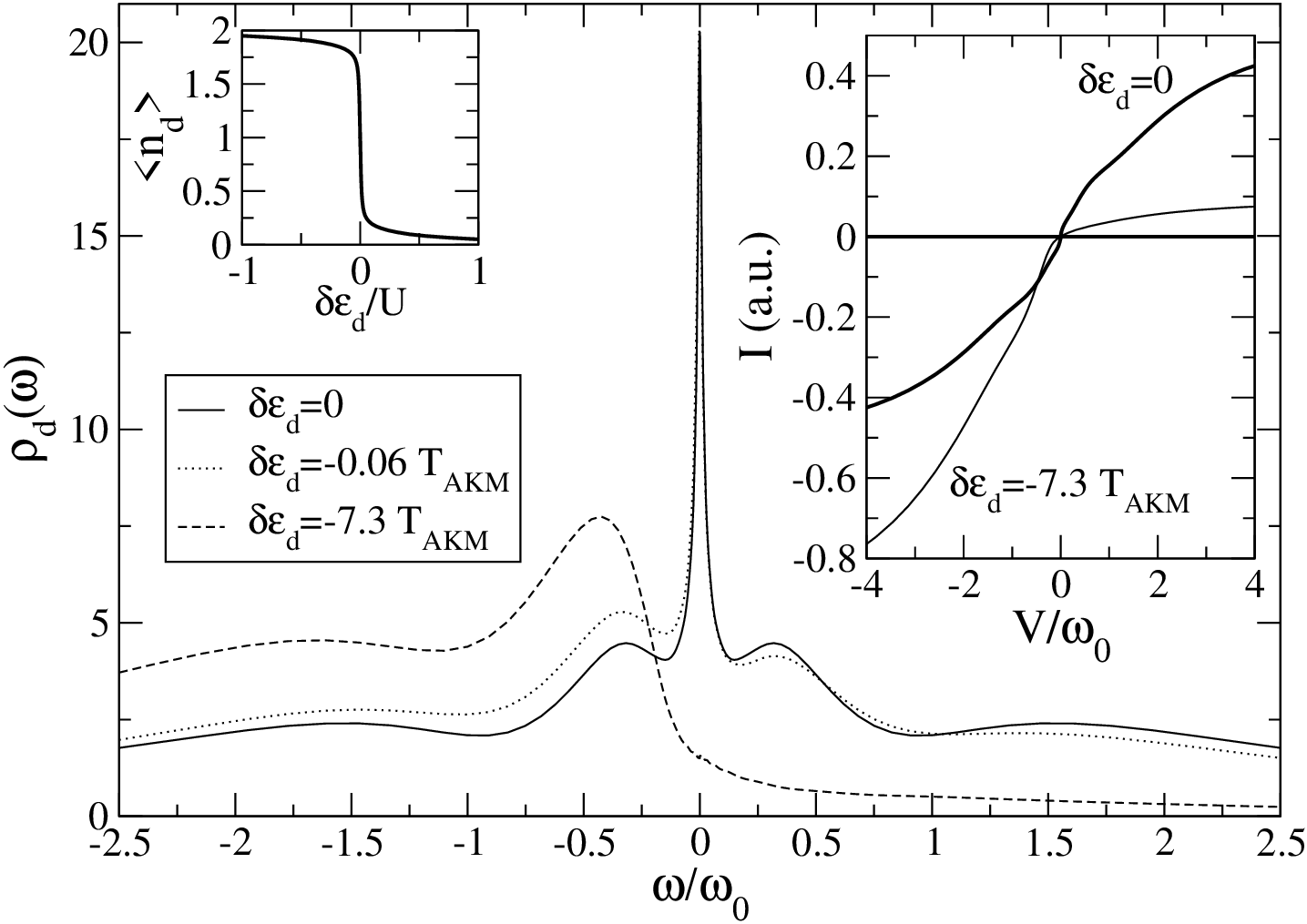}
\caption{Spectral density $\rho_d$  for different values
of the asymmetry $\delta\varepsilon_d=\varepsilon_d-\varepsilon^\star_d$
in the strong {\it e-ph} coupling regime ($\lambda=1.2\omega_0$).
Here $T_{AKM}\sim 0.0012$.
Left inset : average charge 
$\langle n_d\rangle$ in the molecule as a function of
the asymmetry. 
Right inset: current-voltage curves showing
rectification in the asymmetric case.
Other parameters as in Fig. \ref{fig:fig1}.}
\label{fig:densesp}
\end{figure}

Figure~\ref{fig:densesp} shows the spectral density
for strong {\it e-ph} coupling and several values of
 $\delta\varepsilon_d\le 0$. 
At the symmetric point there are peaks at the energies of the 
isolated molecule given by Eq.(\ref{eq:enes}). 
There is also a peak at the Fermi level associated with the charge 
Kondo effect. 
Its width is proportional to $T_{AKM}$ [Eq.~(\ref{takm})].
For $\left|\delta\varepsilon_d \right|\ll T_{AKM}$ 
there is little change in the spectral density. 
When the asymmetry increases, spectral weight is 
transferred from positive to negative energies $\omega$ and  
$\rho_d(0)$ is suppressed.
For $\left|\delta\varepsilon_d\right| > T_{AKM}$, the central peak
disappears and almost all the spectral weight is concentrated
in the negative $\omega$ region.
The behavior for positive $\delta\varepsilon_d$ is obtained through  the
transformation $\omega \to -\omega$.
This spectral polarization is the result of the strong
charge polarization induced by the asymmetry. 
It is illustrated in the left inset to Figure \ref{fig:densesp} that 
shows the ground-state charge of the molecule $\langle n_d\rangle$ 
as a function of  $\delta\varepsilon_d$. 
It can be seen that charge fluctuations in the molecule are rapidly 
suppressed for $\left|\delta\varepsilon_d\right|\gtrsim T_{\text AKM}$.

An interesting consequence of this spectral asymmetry is
that the molecular 
device exhibits strong rectifying properties in the strong {\it e-ph}
coupling limit.
The right inset to Fig. \ref{fig:densesp} shows the 
current-voltage ($I-V$) characteristics for the device in this regime
\footnote{
Here we simply assume that
$\rho_d(\varepsilon,V) \sim \rho_d(\varepsilon,V=0)$, assumption 
valid for small $V$ and calculate the current using 
$I\propto \int d\varepsilon
\left[f(\varepsilon+eV)-f(\varepsilon)\right]\rho_d(\varepsilon)$.}.
At the symmetric point, the $I-V$ curve is symmetric around zero
bias $V$ and it exhibits some structures associated to the peaks 
in the spectral density.
In the asymmetric case 
(with $ \delta\varepsilon_d < 0$), the molecule is polarized and 
the $I-V$ curve is asymmetric: the current for negative bias $V$
is much larger than for positive $V$ (the opposite occurs
for $\delta\varepsilon_d>0$).

In summary, 
we have studied the transport properties of a molecular device
including {\it e-e} and {\it e-ph} interactions.
We have shown that at low temperatures and 
weak {\it e-ph} coupling,
the conductance properties of the molecular junction are described by 
 the standard Kondo model with renormalized parameters.
At large {\it e-ph} coupling, new physics appears:
the properties of the device are described in terms of 
a charge analog of the Kondo effect. 
In this regime the transport properties of
the molecular junction are controlled by the large charge
polarizability of the molecule. The latter induces a strong  
spectral polarization that leads to a rectifying behaviour of
the current-voltage characteristics of the molecular junction. 

We thank  C. Castellani, M. Grilli and  M. J. Rozenberg 
 for useful discussions. 

%\bibliography{refs}

\end{document}